\documentclass[%
 reprint,
 amsmath,amssymb,
 aps,
]{revtex4-2}

\usepackage{graphicx}
\usepackage{dcolumn}
\usepackage{bm}

\begin{document}
\title{Role of Disorder in the Intrinsic Orbital Hall Effect}

\author{Ping Tang$^{1}$}
\email{tang.ping.a2@tohoku.ac.jp}
\author{Gerrit E. W. Bauer$^{1,2,3,4}$}
\affiliation{$^1$WPI-AIMR, Tohoku
University, 2-1-1 Katahira, Sendai 980-8577, Japan}
\affiliation{$^2$Institute for Materials Research, Tohoku University,
2-1-1 Katahira, Sendai 980-8577, Japan} 
\affiliation{$^3$Center for
Spintronics Research Network, Tohoku University, Sendai 980-8577, Japan}
\affiliation{$^4$Kavli Institute for Theoretical Sciences, University of the Chinese Academy of Sciences,
Beijing 10090, China}
\date{\today}

\begin{abstract}
We investigate the effect of random defect scattering on the orbital Hall effect by solving a quantum Boltzmann equation. Depending on the specific orbital textures, diffuse scattering by an \emph{arbitrarily} weak disorder can affect and even fully suppress an intrinsic orbital Hall current. From the results for a simple model, we infer that disorder can play an important role in orbitronics in general.
\end{abstract}
\maketitle
\emph{Introduction.---}The orbital angular momentum of an electron is a fundamental degree of freedom, but unlike its intrinsic spin typically quenched in solids by crystal fields \cite{Kittel2004}. Nevertheless, internal or external electric fields can induce nonzero orbital angular momenta into Bloch states, leading to an orbital Hall effect (OHE, not to be confused with the ordinary Hall effect) \cite{Bernevig2005,Kontani2007,Kontani2008,Tanaka2008,Kontani2009,Go2018,Jo2018,Choi2023}, an orbital Rashba-Edelstein effect \cite{Park2011, Park2015,Yoda2015, Go2017,Yoda2018,Salemi2019,Johansson2021,Ding2022, Hamdi2023}, and an orbital torque \cite{Chen2018,Ding2020,Go2020,Zheng2020, Tazaki2020,Kim2021,Lee2021,Hayashi2023}. The intrinsic OHE appears to be similar to the intrinsic spin Hall effect \cite{Sinova2004}, which is caused by spin textures in momentum space in the presence of spin-orbit interaction and broken inversion symmetry. However, the orbital textures in momentum space responsible for the intrinsic OHE exist already in centrosymmetric and non-relativistic systems \cite{Go2018}. The large orbital Hall conductivity (OHC) calculated for a wide range of materials \cite{Kontani2008, Jo2018,Sala2022} and confirmed by experiments \cite{Choi2023} energizes the interest in ``orbitronics" \cite{Bernevig2005,Go2021}. Existing theories are based on the linear-response Kubo formalism that in the absence of disorder leads to a finite ``intrinsic" OHC. Even though the omnipresent disorder is essential for the spin Hall effect \cite{Inoue2004, Khaetskii2006, Sinova2015},  its role in the OHE has not yet been addressed conclusively. Bernevig \emph{et al.} \cite{Bernevig2005} argued that diffuse scattering described by the orbital current vertex correction vanishes by symmetry in p-doped Si. However, analogous to the vanishing vertex correction to the charge current in normal metals \cite{Mahan}, this result is valid only in the limit of short-range scattering potentials. Tanaka \emph{et al.} \cite{Tanaka2008} calculated finite but small corrections to the OHE in a tight-binding model for transition metals, but again only for single-site substitutional disorder.  

In this Letter, we report that the OHC strongly depends on the details of the orbital character and the disorder. It may even vanish identically regardless of the disorder scattering strength. We model the diffuse scattering by the in-scattering contribution to the collision term in the linearized Boltzmann equation, which is equivalent to the vertex correction to the intrinsic Kubo formula \cite{Mahan}. The finite range of the impurity potentials, neglected in previous theories, turns out to be crucial. 

\begin{figure}
\centering
\par
\includegraphics[width=7.6cm]{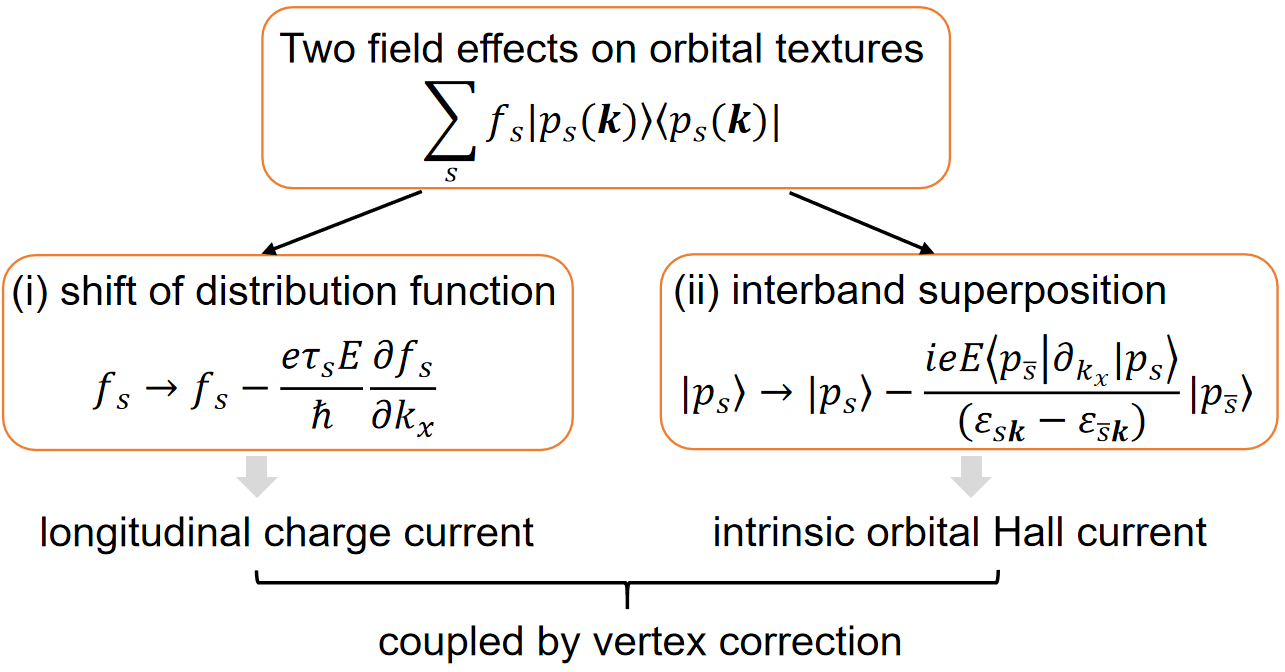}
\newline \caption{An external electric field \textit{E}\textbf{x} affects the orbital texture in a two-band model Eq.~(\ref{Ham2}) with eigenstates $\vert p_{s}(\mathbf{k})\rangle$ and energies $\varepsilon_{s\mathbf{k}}$, where \(s (\overline s) \) are the band indices and \textbf{k} the wave vector, by (i) a shift of the Fermi-Dirac distribution $f_{s}$ of carriers that for a constant relaxation time ($\tau_{s}$) is uniform and (ii) finite matrix elements between unperturbed eigenstates that generate nonzero orbital angular momenta proportional to the ratio of the field to the interband splitting. Diffuse scattering described by the in-scattering term in the collision integral (vertex correction) couples the two mechanisms that ultimately may lead to a vanishing orbital Hall effect for the orbital textures shown in Fig.~\ref{Fig-2}(d) (see text).}%
\label{Fig-1}%
\end{figure}

We illustrate our findings at the hand of a generic two-band $p$-orbital tight-binding model \cite{Go2018}. An applied external electric field has two effects [see Fig.~(\ref{Fig-1})]: it redistributes the electron occupation number near the Fermi level and causes interband mixing. The former generates a charge current relaxed by the disorder scatterings, while the latter emerges from the momentum dependence of the orbital texture eigenstates and corresponds to a transverse orbital angular momentum current. Diffuse scattering mixes the two processes at the Fermi energy and thereby affects the orbital Hall current depending on the nature and strength of the disorder. 

\emph{Generic two-band model.---}We consider a two-dimensional lattice of $p$-orbitals in the $xy$ plane with Hamiltonian
\begin{equation}
\hat{\mathcal{H}}=\sum_{\mathbf{k}}\left(\begin{matrix}
\hat{c}_{p_{x}\mathbf{k}}^{\dagger}  \\
\hat{c}_{p_{y}\mathbf{k}}^{\dagger}
\end{matrix}\right)^{T}\left(\begin{matrix}
\varepsilon_{p_{x}\mathbf{k}} & H_{p_{x}p_{y}}  \\
H_{p_{x}p_{y}}^{\ast} &\varepsilon_{p_{y}\mathbf{k}}
\end{matrix}\right)\left(\begin{matrix}
\hat{c}_{p_{x}\mathbf{k}}  \\
\hat{c}_{p_{y}\mathbf{k}}
\end{matrix}\right)
\end{equation}
where $\hat{c}_{p_{x(y)}\mathbf{k}}$ and $\hat{c}_{p_{x(y)}\mathbf{k}}^{\dagger}$ are, respectively, the annihilation and creation operators of Bloch states $\vert\psi_{p_{x(y)}\mathbf{k}}\rangle=(1/\sqrt{N})\sum_{i}e^{i\mathbf{k}\cdot\mathbf{R}_{i}}\vert\varphi_{p_{x(y)}\mathbf{R}_{i}}\rangle$ formed by atomic-like wavefunctions $\vert\varphi_{p_{x(y)}\mathbf{R}_{i}}\rangle$ centered at $\mathbf{R}_{i}$. Here $\mathbf{k}=k(\cos\phi_{\mathbf{k}}\mathbf{x}+\sin\phi_{\mathbf{k}}\mathbf{y})$ is a crystal wave vector with azimuth angle $\phi_{\mathbf{k}}$, $N$ is the number of unit cells, and $\varepsilon_{p_{x(y)}\mathbf{k}}$ is the associated energy dispersion. $H_{p_{x}p_{y}}(\mathbf{k})$ is an inter-orbital matrix element that is finite in a triangular lattice or induced by $sp$ hybridization in a cubic one \cite{Han2023}. The $p_{z}$-orbital band may be disregarded since it decouples from the $p_{x}$ and $p_{y}$ orbitals due to the 2D mirror symmetry. In the basis of ($\vert p_{x}\rangle$, $\vert p_{y}\rangle$), where $\vert p_{x(y)}\rangle=e^{-i\mathbf{k}\cdot\mathbf{r}}\vert\psi_{p_{x(y)}\mathbf{k}}\rangle$ are the atomic cell-periodic Bloch functions, the above two-band Hamiltonian reads in momentum space 
\begin{equation}
\hat{\mathcal{H}}(\mathbf{k})=d_{0}(\mathbf{k})\hat{1}+\mathbf{d}(\mathbf{k})\cdot\hat{\boldsymbol{l}}\label{Ham2}
\end{equation}
where $\hat{1}$ is the $2\times 2$ identity matrix, and $\boldsymbol{l}$ the vector of pseudospin Pauli matrices: $\hat{l}_{x}=\left(\begin{matrix}
0 & 1\\
1 &0
\end{matrix}\right)$, $\hat{l}_{y}=\left(\begin{matrix}
0 & -i\\
i &0
\end{matrix}\right)$ and $\hat{l}_{z}=\left(\begin{matrix}
1 & 0\\
0 &-1
\end{matrix}\right)$, with $d_{0}=(\varepsilon_{p_{x}\mathbf{k}}+\varepsilon_{p_{y}\mathbf{k}})/2$, $d_{x}=\text{Re}[H_{p_{x}p_{y}}]$, $d_{y}=-\text{Im}[H_{p_{x}p_{y}}]$ and $d_{z}=(\varepsilon_{p_{x}\mathbf{k}}-\varepsilon_{p_{y}\mathbf{k}})/2$. In a centrosymmetric system, $H_{p_{x}p_{y}}=H_{p_{x}p_{y}}^{\ast}$ \cite{Hpxpy}, such that $d_{y}=0$ and $\mathbf{d}$ lies within the $xz$ plane. The eigenvalues of Eq.~(\ref{Ham2}) are $\varepsilon_{\pm\mathbf{k}}=d_{0}\pm \vert\mathbf{d}\vert$, with the corresponding eigenstates being $\vert p_{+}(\mathbf{k})\rangle=\cos\frac{\Theta_{\mathbf{k}}}{2}\vert p_{x}\rangle+\sin\frac{\Theta_{\mathbf{k}}}{2}\vert p_{y}\rangle$ and  $\vert p_{-}(\mathbf{k})\rangle=\sin\frac{\Theta_{\mathbf{k}}}{2}\vert p_{x}\rangle-\cos\frac{\Theta_{\mathbf{k}}}{2}\vert p_{y}\rangle$, where $\Theta_{\mathbf{k}}=\arctan{(d_{x}}/{d_{z}})$ is the angle between $\mathbf{d}(\mathbf{k})$ and the $z$ direction. $H_{p_{x}p_{y}}(\mathbf{k})$ causes the orbital textures $\vert p_{\pm}(\mathbf{k})\rangle$ in momentum space, which is the root of the intrinsic orbital Hall effect \cite{Go2018}. 

We address the Hall current of orbital angular momentum along the \textit{z}-direction in the $y$-direction \(J_{y}^{z}\) under an external electric field $E\mathbf{x}$:
\begin{align}
J_{y}^{z}=& \frac{1}{2}\int\frac{d\mathbf{k}}{(2\pi)^2}\text{Tr}\left[\hat{\rho}\left(\hat{v}_{y} \hat{L}_{z} +\hat{L}_{z}\hat{v}_{y}\right)\right]\nonumber\\
=&-i \int\frac{d\mathbf{k}}{(2\pi)^2}  \frac{\partial d_{0}}{\partial k_{y}} \left(\rho_{+-}-\rho_{-+}\right) \label{OHC}
\end{align}
where $\hat{v}_{y}=\partial \hat{\mathcal{H}}(\mathbf{k})/\partial (\hbar k_{y}) $ is the $y$ (transverse)-component of the group velocity operator  \cite{Notepxpy}, $\hat{L}_{z}$ the $z$-component of the orbital angular momentum operator with matrix elements $\langle p_{x}\vert \hat{L}_{z}\vert p_{y}\rangle=-i\hbar$ for the atomic wave function on the same site (zero otherwise) \cite{Bernevig2005,Kontani2007,Kontani2008,Tanaka2008,Kontani2009,Go2018,Jo2018}, and $\hat{\rho}=\sum_{s, s^{\prime}} \rho_{ss^{\prime}}\vert p_{s}(\mathbf{k})\rangle\langle p_{s^{\prime}}(\mathbf{k})\vert$ the density matrix in the orbital-texture basis with $s,s^{\prime}=\pm$. In the second equality of Eq.~(\ref{OHC}) only the average group velocity $\partial d_{0}/(\hbar\partial k_{y})$ contributes to the orbital Hall current since $\langle p_{s}\vert \hat{L}_{z}\vert p_{s}\rangle=0$. The OHE therefore emerges from the non-diagonal components of the density matrix or a coherent superposition of the \(\vert p_{x}\rangle\)- and \(\vert p_{y}\rangle\)-bands in the presence of the electric field. Our two-band model can be applied to the $d$-orbital OHE of transition metals by substituting the $p_{x}$ and $p_{y}$ orbitals by $d_{yz}$ and $d_{zx}$ or $d_{xy}$ and $d_{x^{2}-y^2}$ pairs \cite{Kontani2009} 

The Boltzmann equation of density matrix that includes the quantum coherence between the two bands reads \cite{Dyakonov1984,Khaetskii2006,Culcer2017}
\begin{equation}
\frac{\partial \hat{\rho}(\mathbf{k})}{\partial t}+\frac{eE \mathbf{x}}{\hbar}\cdot\frac{\partial \hat{\rho}^{(eq)}}{\partial\mathbf{k}}+\frac{i}{\hbar}\left[\hat{\mathcal{H}}(\mathbf{k}), \hat{\rho}(\mathbf{k})\right]=\left.\frac{\partial\hat{\rho}(\mathbf{k})}{\partial t}\right\vert_{\text{col}}, \label{KEQ}
\end{equation}
where $e$ is the elementary charge of the carriers, \textit{i.e.}, $e<0$ ($e>0$) for electrons (holes), and $[\cdots]$ is a commutator. The equilibrium density matrix $\hat{\rho}^{(eq)}=\sum_{s}f_{s}\vert p_{s}\rangle\langle p_{s}\vert$, where $f_{s}(\varepsilon_{s\mathbf{k}})$ is the Fermi-Dirac distribution function of electrons with energy $\varepsilon_{s\mathbf{k}}$, is diagonal in the orbital-texture basis $\vert p_{\pm}(\mathbf{k})\rangle$. In its derivative
\begin{equation}
\frac{\partial\hat{\rho}^{(eq)}}{\partial\mathbf{k}}=\sum_{s}\frac{\partial f_{s}}{\partial\mathbf{k}} \vert p_{s}\rangle\langle p_{s}\vert+f_{s}\left(\frac{\partial\vert p_{s}\rangle}{\partial\mathbf{k}}\langle p_{s}\vert+\text{H.c.}\right)
\end{equation}
the momentum dependent orbital texture introduces off-diagonal components by $\partial\vert p_{s}\rangle/\partial\mathbf{k}=(\bar{s}/2)(\partial\Theta_{\mathbf{k}}/\partial \mathbf{k})\vert p_{\bar{s}}\rangle$ \cite{Notepxpy}, where $\bar{s}=-s$. Assuming that $d_{x}(\mathbf{k})$ and $d_{z}(\mathbf{k})$ are homogeneous functions of the same degree, $\partial\Theta_{\mathbf{k}}/\partial\mathbf{k}=(\nu/k)(-\sin\phi_{\mathbf{k}}\mathbf{x}+ \cos\phi_{\mathbf{k}}\mathbf{y})$, where $\nu =\partial\Theta_{\mathbf{k}}/\partial \phi_{\mathbf{k}}$ is constant by the azimuthal symmetry and acts like a ``winding" number.

Potential scattering causes the collision term on the right-hand side of Eq.~(\ref{KEQ}). In the texture basis and after ensemble average over the disorder \cite{Khaetskii2006,Culcer2017}
\begin{align}
\left.\frac{\partial\rho_{ss^{\prime}}(\mathbf{k})}{\partial t}\right\vert_{\text{col}} &=\frac{\pi}{\hbar}\int\frac{d\mathbf{k}^{\prime}}{(2\pi)^2}\sum_{s_{1},s_{1}^{\prime}}[\delta (\varepsilon_{s_{1}\mathbf{k}^{\prime}}-\varepsilon_{s\mathbf{k}})\nonumber\\
+\delta(\varepsilon_{s_{1}^{\prime}\mathbf{k}^{\prime}}&-\varepsilon_{s^{\prime}\mathbf{k}})]\mathcal{M}_{s_{1} s_{1}^{\prime}}^{ss^{\prime}} \rho_{s_{1}s_{1}^{\prime}}(\mathbf{k}^{\prime})-\delta(\varepsilon_{s_{1}^{\prime}\mathbf{k}^{\prime}}-\varepsilon_{s_{1}\mathbf{k}})\nonumber\\\times
[\mathcal{M}_{s_{1}^{\prime}s_{1}^{\prime}}^{ss_{1}}&\rho_{s_{1}s^{\prime}}(\mathbf{k})+\rho_{ss_{1}}(\mathbf{k})\mathcal{M}_{s_{1}^{\prime}s_{1}^{\prime}}^{s_{1}s^{\prime}}]\}.\label{scat}
\end{align}
The first integrand represents the in-scattering contribution that in the diagrammatic language corresponds to the vertex correction, while the second one is the out-scattering term or self-energy. $\mathcal{M}_{s_{1} s_{1}^{\prime}}^{ss^{\prime}}=\overline{\langle\psi_{p_{s}\mathbf{k}}\vert \hat{U}\vert\psi_{p_{s_{1}}\mathbf{k}^{\prime}}\rangle \langle \psi_{p_{s_{1}^{\prime}}\mathbf{k}^{\prime}}\vert \hat{U}\vert \psi_{p_{s^{\prime}}\mathbf{k}}\rangle}$ are the scattering matrix elements in the Born approximation, where the overline implies the average over disorder configurations. Here we adopt a substitutional disorder model in real space $\overline{\langle \varphi_{p_{l}\mathbf{R}_{i}}\vert \hat{U}\vert \varphi_{p_{l_{1}}\mathbf{R}_{i_{1}}}\rangle \langle \varphi_{p_{l_{1}^{\prime}}\mathbf{R}_{i_{1}^{\prime}}}\vert \hat{U}\vert \varphi_{p_{l^{\prime}}\mathbf{R}_{i^{\prime}}}\rangle}=n_{\mathrm{imp}} \overline{V_{\mathbf{R}_{i}}V_{\mathbf{R}_{i^{\prime}}}}\delta_{ll_{1}} \delta_{l_{1}^{\prime}l^{\prime}} \delta_{ii_{1}}\delta_{i_{1}^{\prime}i^{\prime}}
$ \cite{Culcer2017}, where $p_{l}$ denotes the atomic orbitals $p_{x}$ and $p_{y}$, while $n_{\mathrm{imp}}$ and $\overline{V_{\mathbf{R}_{i}}V_{\mathbf{R}_{i^{\prime}}}}$ represent the disorder density and distance correlation function between the $i$-th and $i^{\prime}$-th sites, respectively. In a continuum model, this corresponds to randomly distributed potential scatters $\hat{U} =\sum_{i} V(\mathbf{r}-\mathbf{r}_{i})$, where the sum is over the impurity position $\mathbf{r}_{i}$. This leads to $\mathcal{M}_{s_{1} s_{1}^{\prime}}^{ss^{\prime}}=n_{\mathrm{imp}}\vert V_{\mathbf{k}\mathbf{k}^{\prime}}\vert^{2}\langle p_{s}(\mathbf{k})\vert p_{s_{1}}(\mathbf{k}^{\prime})\rangle\langle p_{s_{1}^{\prime}}(\mathbf{k}^{\prime})\vert p_{s^{\prime}}(\mathbf{k})\rangle$, where $V_{\mathbf{k}\mathbf{k}^{\prime}}$ is  the Fourier transform of the impurity potential and/or their position correlation function; the inner products between Bloch states are $\langle p_{s}(\mathbf{k})\vert p_{s}(\mathbf{k}^{\prime})\rangle=\cos\frac{\Theta_{\mathbf{k}\mathbf{k}^{\prime}}}{2}$,  $\langle p_{s}(\mathbf{k})\vert p_{\bar{s}}(\mathbf{k}^{\prime})\rangle=s\sin\frac{\Theta_{\mathbf{k}\mathbf{k}^{\prime}}}{2}$, and $\Theta_{\mathbf{k}\mathbf{k}^{\prime}}=\Theta_{\mathbf{k}^{\prime}}-\Theta_{\mathbf{k}}$.

\emph{Results.---}When $\varepsilon_{s\mathbf{k}}=\varepsilon_{sk}$ we can solve Eq.~(\ref{KEQ}) analytically by the ansatz
\begin{align}
\hat{\rho}(\mathbf{k})-\hat{\rho}^{(eq)}(\mathbf{k})&=\sum_{s}\cos\phi_{\mathbf{k}} g_{s}(k) \vert p_{s}(\mathbf{k})\rangle\langle p_{s}(\mathbf{k})\vert\nonumber\\
&+\sum_{s}\sin\phi_{\mathbf{k}}g_{s\bar{s}}(k)  \vert p_{s}(\mathbf{k})\rangle\langle p_{\bar{s}}(\mathbf{k})\vert\label{ansat}
\end{align}
where $g_{s}(k)$ and $g_{s\bar{s}}(k)$ are the diagonal and off-diagonal elements of the out-of-equilibrium part, respectively. 

\begin{figure}
\centering
\par
\includegraphics[width=6.6cm]{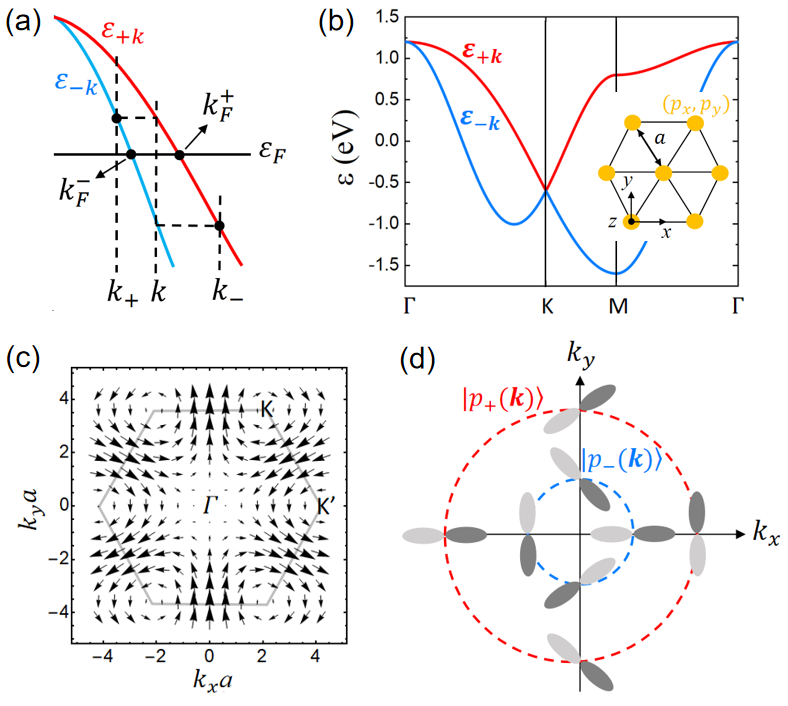}
\newline \caption{(a) Schematics for $k_{\pm}$ and $k_{F}^{\pm}$ defined in the text. (b) Band structures for 2D $p$-orbital triangular lattice of spacing $a$ and (c) the associated textures of the vector $\mathbf{d}(\mathbf{k})=(d_{x}, d_{z})$ plotted in the Brillouin zone for $\epsilon_{p}=0$, $t_{p\sigma}=0.5\,$eV and $t_{p\pi}=-0.1\,$eV. The thickness of the arrows is proportional to $\vert \mathbf{d}(\mathbf{k})\vert$. (d) The orbital textures close to the K-point.}%
\label{Fig-2}%
\end{figure}

In the steady state $\partial\hat{\rho}/\partial t=0$, $g_{s}(k)$ describes a shift of the entire equilibrium distribution in momentum space that is proportional to the field and the momentum relaxation time $\tau_{s}(k)$ as defined below, while the off-diagonal $g_{s\bar{s}}(k)$ that represents interband supposition and scales like $E/\vert \varepsilon_{+k}-\varepsilon_{-k}\vert$ [see Fig.~\ref{Fig-1}]. The former determines the longitudinal charge current, while the latter leads to the transverse orbital Hall current. In the weak scattering limit $(\hbar/\tau_{s})\ll \vert\varepsilon_{+k}-\varepsilon_{-k}\vert$ the intrinsic OHE dominates and $g_{s\bar{s}} \ll g_{s}$ may be dropped in the collision term Eq.~(\ref{scat}). By substituting Eq.~(\ref{ansat}) into Eq.~(\ref{KEQ}), we find
\begin{align}
&\frac{eE}{\hbar}\frac{\partial f_{s}}{\partial k}=-\frac{g_{s}(k)}{\tau_{s}(k)} +c_{1}(k,k_{s})\frac{k_{s} g_{\bar{s}}(k_{s})}{v_{\bar{s}}(k_{s})}\label{gss}\\
&\frac{ \nu eE}{\hbar k}(f_{s}-f_{\bar{s}})+\frac{2si}{\hbar}\left(\varepsilon_{sk}-\varepsilon_{\bar{s}k}\right)g_{s\bar{s}}(k)
 =c_{2}(k,k_{\bar{s}})\frac{k_{\bar{s}}g_{s}(k_{\bar{s}})}{v_{s}(k_{\bar{s}})}\nonumber\\
 &-c_{2}(k,k_{s})\frac{k_{s}g_{\bar{s}}(k_{s})}{v_{\bar{s}}(k_{s})}+ c_{2}(k,k)\left[\frac{kg_{s}(k)}{v_{s}(k)}-\frac{ kg_{\bar{s}}(k)}{v_{\bar{s}}(k)}\right],\label{goff}
   \end{align}
where $\tau_{s}^{-1}(k)=\frac{ka(k,k)}{v_{s}(k)} + \frac{k_{s}b(k,k_{s})}{v_{\bar{s}}(k_{s})}$ is the momentum relaxation rate of an electron in band $s$ and with momentum $k$, and the $c_{1(2)}(k,k^{\prime})$ represent the in-scattering that mixes the $g_{s}$ and $g_{\bar{s}}$ ($g_{s\bar{s}}$) components of the density matrix, with
\begin{subequations}
\begin{align}
a(k,k^{\prime})=&\frac{n_{\text{imp}}}{\hbar^{2}}\int \frac{d\phi_{\mathbf{k}^{\prime}}}{2\pi}\vert V_{\mathbf{k}\mathbf{k}^{\prime}}\vert^{2}\cos^{2}\frac{\Theta_{\mathbf{k}\mathbf{k}^{\prime}}}{2}(1-\cos\phi_{\mathbf{k}\mathbf{k}^{\prime}}),\nonumber\\
b(k,k^{\prime})=&\frac{n_{\text{imp}}}{\hbar^{2}}\int \frac{d\phi_{\mathbf{k}^{\prime}}}{2\pi} \vert V_{\mathbf{k}\mathbf{k}^{\prime}}\vert^{2}\sin^{2}\frac{\Theta_{\mathbf{k}\mathbf{k}^{\prime}}}{2},\nonumber\\
c_{1}(k,k^{\prime})=&\frac{n_{\text{imp}}}{\hbar^{2}}\int\frac{d\phi_{\mathbf{k}^{\prime}}}{2\pi}\vert V_{\mathbf{k}\mathbf{k}^{\prime}}\vert^{2}\sin^{2}\frac{\Theta_{\mathbf{k}\mathbf{k}^{\prime}}}{2}\cos\phi_{\mathbf{k}\mathbf{k}^{\prime}},\nonumber\\
c_{2}(k,k^{\prime})=&\frac{n_{\text{imp}}}{\hbar^{2}}\int \frac{d\phi_{\mathbf{k}^{\prime}}}{2\pi} \vert V_{\mathbf{k}\mathbf{k}^{\prime}}\vert^{2}\sin\frac{\Theta_{\mathbf{k}\mathbf{k}^{\prime}}}{2}\cos\frac{\Theta_{\mathbf{k}\mathbf{k}^{\prime}}}{2}\sin\phi_{\mathbf{k}\mathbf{k}^{\prime}}\nonumber.
\end{align}  
\end{subequations}
$c_{2}(k,k^{\prime})$ couples the transverse orbital Hall current to the charge current. $k_{s}$ and $k_{\bar{s}}$ are, respectively, the moduli of the wave vectors of the final state of the energy-conserving \emph{interband} scattering events from an initial state $k$ into the band $s$ and $\bar{s}$ with $\varepsilon_{\bar{s}k_{s}}=\varepsilon_{sk}$ and $\varepsilon_{sk_{\bar{s}}}=\varepsilon_{\bar{s}k}$, as illustrated in Fig.~\ref{Fig-2}(a). Since the relaxation terms associated with $g_{s\bar{s}}$ are small for weak impurity scattering, the in-scattering (vertex correction) term dominates the right-hand side of Eq.~(\ref{goff}).

Eqs.~(\ref{gss}) and (\ref{goff}) connect the non-equilibrium distributions at the initial $k$ with the ones at the scattered $k_{s}$ and $k_{\bar{s}}$. By taking $s=\bar{s}$ and $k=k_{s}$ in Eq.~(\ref{gss}), we obtain another two equations that combined with Eq.~(\ref{gss}) give
\begin{align}
g_{s}(k)=-\frac{e\widetilde{\tau}_{s}(k) E}{\hbar}\frac{\partial f_{s}(\varepsilon_{sk})}{\partial k} 
\end{align}
where we introduced a transport relaxation time $\widetilde{\tau}_{s}(k)$ that is modified by the interband $c_{1}(k,k_{s})$-vertex correction:
\begin{align}\widetilde{\tau}_{s}(k) = 
\frac{\frac{1}{\tau_{\bar{s}}(k_{s})} 
+\frac{k_{s}c_{1}(k,k_{s})}{v_{s}(k)}}{\frac{1}{\tau_{\bar{s}}(k_{s})\tau_{s}(k)}-\frac{kk_{s}c_{1}(k,k_{s})c_{1}(k_{s},k)}{v_{s}(k)v_{\bar{s}}(k_{s})}}
\end{align}
and we used the relation $\partial f_{\bar{s}}/\partial k_{s}=(\partial f_{s}/\partial k) v_{\bar{s}}(k_{s})/v_{s}(k)$. According to Eq.~(\ref{goff}), the off-diagonal elements read
\begin{align}
g_{s\bar{s}}(k)=&\frac{ i eE}{2(\varepsilon_{+k}-\varepsilon_{-k})}\left\{\frac{\nu }{ k}\left(f_{s}-f_{\bar{s}}\right)+\left[c_{2}(k,k)\frac{k\widetilde{\tau}_{s}(k)}{v_{s}(k)}\right.\right.\nonumber\\&
\left.\left.-c_{2}(k,k_{s})\frac{k_{s}\widetilde{\tau}_{\bar{s}}(k_{s})}{v_{s}(k)}\right] \frac{\partial f_{s}}{\partial k}-\left[c_{2}(k,k)\frac{k\widetilde{\tau}_{\bar{s}}(k)}{v_{\bar{s}}(k)}\right.\right.\nonumber\\&\left.\left.-c_{2}(k,k_{\bar{s}})\frac{k_{\bar{s}}\widetilde{\tau}_{s}(k_{\bar{s}})}{v_{\bar{s}}(k)}\right]\frac{\partial f_{\bar{s}}}{\partial k}\right\}\label{Off2}
\end{align}
which leads to
\begin{align}
\sigma_{\text{OHC}}&\equiv \frac{J_{y}^{z}}{E}= \frac{ e}{8\pi}\sum_{s=\pm}s\left\{\int \frac{ \nu f_{s}}{\vert \mathbf{d}(k)\vert}\frac{\partial d_{0}}{\partial k} dk-\frac{k_{F}^{s}}{\vert\mathbf{d}(k_{F}^{s})\vert}\frac{\partial d_{0}}{\partial k_{F}^{s}}\right.\nonumber\\
&\left.\times\left[c_{2}(k_{F}^{s},k_{F}^{s})\frac{k_{F}^{s}\widetilde{\tau}_{s}(\varepsilon_{F})}{v_{F}^{s}}-c_{2}(k_{F}^{s},k_{F}^{\bar{s}})\frac{k_{F}^{\bar{s}}\widetilde{\tau}_{\bar{s}}(\varepsilon_{F})}{v_{F}^{s}}\right]\right\} \label{OHE}
\end{align}
where $k_{F}^{s}$, $v_{F}^{s}$, and $\widetilde{\tau}_{s}(\varepsilon_{F})$ are the wave vector, group velocity, and transport relaxation time of band \textit{s} at the Fermi level, respectively. 

The two bands contribute to the OHC in the presence of disorder with opposite signs. The first term in the curly bracket on the right-hand side of Eq.~(\ref{OHE}) represents the ``intrinsic" contribution from field-induced interband mixing and agrees with the Kubo formalism without the disorder. The second term is the (in-scattering) vertex correction that includes both intraband [$c_{2}(k_{F}^{s},k_{F}^{s})$] and interband [$c_{2}(k_{F}^{s},k_{F}^{\bar{s}})$] contributions. Since $c_{2}(k,k^{\prime})\propto n_{\mathrm{imp}}\vert V_{\mathbf{k}\mathbf{k}^{\prime}} \vert^{2}$, the latter does \emph{not} depend on the $n_{\text{imp}}$ and scattering strength and thereby exists for an \emph{arbitrarily} weak impurity. Thus this term cannot be disentangled experimentally from the intrinsic OHC, quite analogous to the side jump mechanism in the spin Hall effect \cite{Sinova2015}. We note that the orbital spin Hall angle between the orbital and charge currents  \(\theta_{\mathrm{OHE}}=J_{y}^{z}/J_c\) vanishes in the ballistic limit, implying an inefficient generation of the orbital Hall current in a good metal. In contrast to the statements in Refs.~\cite{Bernevig2005,Tanaka2008,Go2018}, the vertex correction does not have to vanish or be negligibly small. Instead, we show now that it \textit{can} be very significant for a generic model. 

\emph{2D triangular lattice.---}We focus here on a 2D triangular (Bravais) lattice of spacing $a$, as illustrated in Fig.\ref{Fig-2}(b). While the calculations are more tedious for the honeycomb (graphene) band structure, we expect results to be similar. The tight-binding model of \((p_x,p_y)\) orbitals gives $d_{0}=\epsilon_{p}+(t_{p\sigma}+t_{p\pi})(\cos k_{x}a+2\cos\frac{k_{x}a}{2}\cos\frac{\sqrt{3}k_{y}a}{2})$, $d_{x}=\sqrt{3}(t_{p\pi}-t_{p\sigma})\sin\frac{k_{x}a}{2}\sin\frac{\sqrt{3}k_{y}a}{2}$, $d_{y}=0$ and $d_{z}=(t_{p\sigma}-t_{p\pi})(\cos k_{x}a-\cos\frac{k_{x}a}{2}\cos\frac{\sqrt{3}k_{y}a}{2})$ \cite{Han2023}, where $\epsilon_{p}$ is the degenerate on-site energy of the $p$-orbitals, and $t_{p\sigma}$ and $t_{p\pi}$ are the Slater-Koster hopping amplitudes for, respectively, $\sigma$ and $\pi$ bonds with $|t_{p\sigma}|>|t_{p\pi}\vert$ in general. Fig. \ref{Fig-2}(b) plots the vector $\mathbf{d}=(d_{x}, d_{z})$ in the Brillouin zone. Below we discuss the OHE around some high-symmetry points.

\emph{(i) Near the $\Gamma$ point} and to leading order in $k$, $d_{0}=-\frac{3a^{2}}{4}(t_{p\sigma}+t_{p\pi})k^{2}+\text{const.}$, $d_{x}=\frac{3a^{2}}{4}(t_{p\pi}-t_{p\sigma})k_{x}k_{y}$, and $d_{z}=\frac{3a^{2}}{8}(t_{p\sigma}-t_{p\pi}) (k_{y}^{2}-k_{x}^{2})$, which corresponds to $\Theta_{\mathbf{k}}=2\phi_{\mathbf{k}}+\pi$ (i.e., $\nu=2$), and $\vert p_{+}(\mathbf{k})\rangle$ and $\vert p_{-}(\mathbf{k})\rangle$ reduce to the common tangential and radial orbital states $\vert p_{t}(\mathbf{k})\rangle=-\sin\phi_{\mathbf{k}}\vert p_{x}\rangle+\cos\phi_{\mathbf{k}}\vert p_{y}\rangle$ and $\vert p_{r}(\mathbf{k})\rangle=\cos\phi_{\mathbf{k}}\vert p_{x}\rangle+\sin\phi_{\mathbf{k}}\vert p_{y}\rangle$, respectively, with $\varepsilon_{\pm k}=-\frac{3a^{2}}{4}(t_{p\sigma}+t_{p\pi})k^{2}\pm \frac{3a^{2}}{8}\vert t_{p\sigma}-t_{p\pi}\vert k^{2}$. In the limiting case of $\delta$-correlated (point-like) scatters with $V_{\mathbf{k}\mathbf{k}^{\prime}}=\text{const.}$, the vertex correction $c_{2}(k,k^{\prime})\propto\int d\phi_{\mathbf{k}^{\prime}}\sin^{2}\phi_{\mathbf{k}\mathbf{k}^{\prime}}\cos\phi_{\mathbf{k}\mathbf{k}^{\prime}}$ vanishes. When the Fermi level crosses both bands (i.e., $t_{p\sigma}>-3t_{p\pi}$), Eq.~(\ref{OHE}) then reduces to the intrinsic value
 \begin{equation}
  \sigma_{\text{int}}=\frac{e}{2\pi}\frac{m_{+}^{\ast}+m_{-}^{\ast}}{m_{-}^{\ast}-m_{+}^{\ast}} \ln\frac{k_{F}^{+}}{k_{F}^{-}}. \label{intrsigma}
 \end{equation}
The effective masses $m_{+}^{\ast}=-\frac{4\hbar^{2}}{3a^{2}(t_{p\sigma}+3t_{p\pi})}$ and $m_{-}^{\ast}=-\frac{4\hbar^{2}}{3a^{2}(3t_{p\sigma}+t_{p\pi})}$ of the $\pm$ bands are negative since the \(\Gamma\) point is an energy maximum. When $t_{p\sigma}\leq-3t_{p\pi}$ the effective mass of the upper band changes sign and only one Fermi surface exists,  leading to a logarithmic divergence of the intrinsic part of Eq.~(\ref{OHE}) \cite{Notediverg}, illustrating the sensitivity of the OHE to the band structure. 
  
Analogous to charge transport in metals, the vertex correction vanishes only for \(\delta\)-function potentials \cite{Mahan}. Here we show that for any \emph{finite}-range potentials $c_{2}(k,k^{\prime})\neq 0$ may dramatically modify the OHC. We illustrate this general conclusion by charged impurities with a statically screened 2D Coulomb potential $V_{\mathbf{k}\mathbf{k}^{\prime}}=V_{0}/\sqrt{1+\vert\mathbf{k}-\mathbf{k}^{\prime}\vert^2\lambda^{2}}$, where the screening length $\lambda$ measures the reach (or correlation distance) of the impurity potential, e.g., $\lambda= 0$ corresponds to $\delta$-correlated scatters. For a single band at the Fermi level or $\vert k_{F}^{+}-k_{F}^{-}\vert \gg \lambda^{-1}$, we may disregard the interband scatterings. In the former limit, the vertex correction contributes to the OHC by
 \begin{align}
\sigma_{\text{vert}}^{s}&=-\frac{se}{4\pi}\frac{m_{+}^{\ast}+m_{-}^{\ast}}{m_{-}^{\ast}-m_{+}^{\ast}}\frac{\int d\phi_{\mathbf{k}^{\prime}}\vert V_{\mathbf{k}\mathbf{k}^{\prime}}\vert^{2}\sin^{2}\phi_{\mathbf{k}\mathbf{k}^{\prime}}\cos\phi_{\mathbf{k}\mathbf{k}^{\prime}}}{\int d\phi_{\mathbf{k}^{\prime}} \vert V_{\mathbf{k}\mathbf{k}^{\prime}}\vert^{2} \cos^{2}\phi_{\mathbf{k}\mathbf{k}^{\prime}}(1-\cos\phi_{\mathbf{k}\mathbf{k}^{\prime}})}\nonumber\\
&\overset{k_{F}^{s}\lambda\gg1}{=}-\frac{se}{4\pi}\frac{m_{+}^{\ast}+m_{-}^{\ast}}{m_{-}^{\ast}-m_{+}^{\ast}}\left[1-\frac{2\sqrt{2}}{k_{F}^{s}\lambda} \right].\label{Gamvert}
\end{align}
Remarkably, the Fermi wave numbers and screening length do not contribute to an OHC that is of the same order as the intrinsic one. When two bands reside at the Fermi level (but still $\vert k_{F}^{+}-k_{F}^{-}\vert \gg \lambda^{-1}$)
\begin{equation}
 \sigma_{\text{vert}}=\sum_{s}\sigma_{\text{vert}}^{s}=-\frac{\sqrt{2}e}{2\pi}\frac{m_{+}^{\ast}+m_{-}^{\ast}}{m_{-}^{\ast}-m_{+}^{\ast}}\left[\frac{1}{k_{F}^{-}\lambda}-\frac{1}{k_{F}^{+}\lambda} \right]
\end{equation}
which gives a minor correction to Eq.~(\ref{intrsigma}), however.  

\emph{(ii) Near the $\mathbf{K}=(\frac{2\pi}{3a},\frac{2\pi}{\sqrt{3}a})$ point }$d_{0}=\frac{3a^{2}}{8}(t_{p\sigma}+t_{p\pi})\tilde{k}^{2}+\text{const.}$, $d_{x}=\frac{3\sqrt{3}a}{4}(t_{p\sigma}-t_{p\pi}) \tilde{k}_{y}$, and $d_{z}=-\frac{3\sqrt{3}a}{4}(t_{p\sigma}-t_{p\pi})\tilde{k}_{x}$, where $\tilde{\mathbf{k}}=\mathbf{k}-\mathbf{K}$ is a small wave vector. This corresponds to $\Theta_{\tilde{\mathbf{k}}}=\pi-\phi_{\tilde{\mathbf{k}}}$ (when $t_{p\sigma}>t_{p\pi}$) and leads to the Rashba-type orbital textures with $\vert p_{+}(\tilde{\mathbf{k}})=\sin\frac{\phi_{\tilde{\mathbf{k}}}}{2}\vert p_{x}\rangle+\cos\frac{\phi_{\tilde{\mathbf{k}}}}{2}\vert p_{y}\rangle$ and $\vert p_{-}(\tilde{\mathbf{k}})\rangle=\cos\frac{\phi_{\tilde{\mathbf{k}}}}{2}\vert p_{x}\rangle-\sin\frac{\phi_{\tilde{\mathbf{k}}}}{2}\vert p_{y}\rangle$ [see Fig.~\ref{Fig-2}(d)]. The two bands form a tilted Dirac cone centered at the $\mathbf{K}$ point. The intrinsic OHC of the $s$ band reads
\begin{equation}
    \sigma_{\text{int}}^{s}=-\frac{se}{8\sqrt{3}\pi} \frac{t_{p\sigma}+t_{p\pi}}{\vert t_{p\sigma}-t_{p\pi}\vert} \tilde{k}_{F}^{s}a \label{intrK}
\end{equation}
where $e<0$ ($e>0$) for the upper (lower) cone. Elastic interband scatterings are forbidden and the contribution from the vertex correction in Eq.~(\ref{OHE}) reduces to 
 \begin{align}
\sigma_{\text{vert}}^{s}=-\frac{se}{8\sqrt{3}\pi}\frac{t_{p\sigma}+t_{p\pi} }{\vert t_{p\sigma}-t_{p\pi} \vert} \frac{c_{2}(\tilde{k}_{F}^{s},\tilde{k}_{F}^{s})}{a(\tilde{k}_{F}^{s},\tilde{k}_{F}^{s})}\tilde{k}_{F}^{s}a\label{Vert}
 \end{align}
By substituting explicit expressions of $c_{2}(\tilde{k}_{F}^{s},\tilde{k}_{F}^{s})$ and $a(\tilde{k}_{F}^{s},\tilde{k}_{F}^{s})$ along with $\Theta_{\tilde{\mathbf{k}}}=\pi-\phi_{\tilde{\mathbf{k}}}$, we find the universal value
\begin{align}
\frac{c_{2}(k_{F}^{s},k_{F}^{s})}{a(k_{F}^{s},k_{F}^{s})}=&-\frac{\frac{1}{2}\int d\phi_{\mathbf{k}^{\prime}}\vert V_{\mathbf{k}\mathbf{k}^{\prime}}\vert^{2}\sin^{2}\phi_{\mathbf{k}\mathbf{k}^{\prime}}}{\int d\phi_{\mathbf{k}^{\prime}} \vert V_{\mathbf{k}\mathbf{k}^{\prime}}\vert^{2} \cos^{2}\frac{\phi_{\mathbf{k}\mathbf{k}^{\prime}}}{2}(1-\cos\phi_{\mathbf{k}\mathbf{k}^{\prime}})}\nonumber\\
&\equiv -1 !
\end{align}
Remarkably, Eq.~(\ref{Vert}) exactly cancels the intrinsic Eq.~(\ref{intrK}), leading to a vanishing OHE, irrespective of the impurity character. In contrast, the vertex correction fully suppresses the spin Hall effect in the Rashba 2DEG only for \(\delta\)-scattering potentials \cite{Sinova2015}. Since the OHE is even under time reversal, the same results hold near $\mathbf{K}^{\prime}=(-\frac{2\pi}{3a},\frac{2\pi}{\sqrt{3}a})$, so the valley degeneracy leads to a factor 2 in the OHC. The OHE should also vanish for Dresselhaus-type orbital textures with $\nu=+1$, since both the intrinsic and the vertex correction change sign. 

\emph{Conclusions.---}We revealed the role of disorder in the orbital Hall effect by solving a quantum Boltzmann equation. While the in-scattering term of the collision integral or vertex correction to the OHE vanishes for the radial and tangential orbital states in the limit of short-range potentials \cite{Bernevig2005,Go2018}, we show that such a conclusion is not possible otherwise. Finite-range scattering potentials may modify the intrinsic OHC by the same order of magnitude, and the OHE in the K-valleys of the triangular lattice model is exactly suppressed irrespective of the scatter strength and properties. While this result does not hold in general, the nature of the vertex correction does not depend on the detailed model Hamiltonian and should be considered important unless proven otherwise \cite{Mahan}.

\emph{Acknowledge.---}We acknowledge support by JSPS KAKENHI Grants No. 19H00645, No. 22H04965 and No. 23K13050.


\begin{thebibliography}{99}
\bibitem{Hpxpy}We disregard on-site hybridization matrix elements between $p_{x}$ and $p_{y}$ orbitals by spin-orbit coupling that would be purely imaginary and $\mathbf{k}$-independent.
\bibitem{Notepxpy}The tight-binding approximation implicates neglect of the momentum derivative of the atomic orbitals $\vert p_{x}\rangle$ and $\vert p_{y}\rangle$ that make up the Bloch wavefunctions.
\bibitem{Notediverg}Such a logarithmic divergence exists only in a 2D system and would be regulated when the third dimension is introduced.
\end{thebibliography}
\end{document}